\documentclass[twocolumn,showpacs,superscriptaddress,amsmath,amssymb,prl]{revtex4}

\usepackage{graphicx}%
\usepackage{dcolumn}%

\begin{document}

\title{Bare electron dispersion from photoemission experiments}

\author{A. A. Kordyuk}
\affiliation{Institute for Solid State Research, IFW-Dresden, Helmholtzstr.~20, D-01069 Dresden, Germany}
\affiliation{Institute of Metal Physics of National Academy of Sciences of Ukraine, 03142 Kyiv, Ukraine}

\author{S. V. Borisenko}
\author{A. Koitzsch}
\author{J. Fink}
\author{M. Knupfer}
\affiliation{Institute for Solid State Research, IFW-Dresden, Helmholtzstr.~20, D-01069 Dresden, Germany}

\author{H. Berger}
\affiliation{Institute of Physics of Complex Matter, EPFL, CH-1015 Lausanne, Switzerland}

\date{April 30, 2004}%

\begin{abstract}
Performing an in-depth analysis of the photoemission spectra along the nodal direction of the high temperature superconductor Bi-2212 we have developed a procedure to determine the underlying electronic structure and established a precise relation of the measured quantities to the real and imaginary parts of the self-energy of electronic excitations. The self-consistency of the procedure with respect to the Kramers-Kronig transformation allows us to draw conclusions on the applicability of the spectral function analysis and on the existence of well defined quasiparticles along the nodal direction even for the underdoped Bi-2212 in the pseudogap state.  
\end{abstract}


\maketitle

With modern angle-resolved photoemission spectroscopy (ARPES) \cite{DamascelliRMP03, Campuzano} one gets a direct snapshot of the density of low energy electronic excited states in the momentum-energy space of 2D solids \cite{VallaSci99, BogdanovPRL00, KaminskiPRL01, BorisenkoPRB2001}. All the interactions of the electrons which are responsible for the unusual normal and superconducting properties of cuprates are encapsulated in such pictures, but are still hard to decipher. One way to take into account these interactions is to consider electronic excitations as quasiparticles which, compared to the non-interacting electrons, are characterized by an additional complex self-energy \cite{AGD}. Extraction of the self-energy from experiment is thus of great importance to check the validity of the quasiparticle concept and understand the nature of interactions involved, but appears to be problematic since the underlying band structure of the bare electrons is a priori unknown. 

One can evaluate the interaction parameters taking the bare band dispersion from band structure calculations \cite{BogdanovPRL00}, however this unavoidably increases the uncertainty of any conclusions on the strength and nature of the interactions involved. A direct determination of the bare band structure from experiment would be much more attractive in this sense. Previously, the bare band dispersion has been assigned to the high binding energy part of the experimental dispersion \cite{JohnsonPRL01}. In Refs.~\onlinecite{KordyukPRB2003, KoitzschPRB2004} we have discussed that the bare Fermi velocity estimated from the nodal ARPES spectra using the Kramers-Kronig (KK) transformation is in reasonable agreement with band structure calculations \cite{AndersenJPCS95, NBS} and with an analysis of the anisotropic plasmon dispersion \cite{NuckerPRB1991}, although it has been pointed out that in order to quantify interaction parameters such as coupling strength \cite{KimPRL2003} or self-energy \cite{Drop} a precise and reliable approach of bare band determination is needed.  

In this Letter we introduce an approach to directly extract the bare band dispersion and the self-energy functions from ARPES spectra. We show that the approach is self-consistent within the highest experimental accuracy available today,
and, applying the procedure to the spectra from the underdoped Bi-2212, we demonstrate the validity of the quasiparticle concept even in the underdoped regime and in the pseudogap state. The detailed description of the procedure as well as its application to other samples of different doping level is given in Ref.~\onlinecite{BareApp}.

\begin{figure}[b!]
\includegraphics[width=8.47cm]{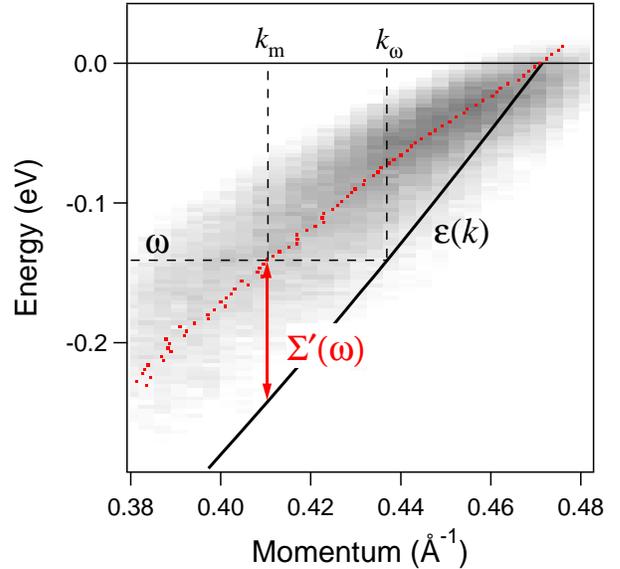}%
\caption{\label{Fig1} Bare band dispersion (solid line) and renormalized dispersion (red points) on top of the spectral weight of interacting electrons. Though intended to be general, this sketch represents the nodal direction of an underdoped Bi-2212.}
\end{figure}

We start with a brief overview of the basics of the nodal spectra analysis within the self-energy approach illustrated in Fig.~\ref{Fig1}. Measuring the photoemission intensity as a function of the kinetic energy and in-plane momentum of outgoing electrons, $I(E_k,\mathbf{k})$, one obtains access to the spectral function of the one electron removal which is supposed to reflect the quasipartical properties of the remaining photohole: its effective mass and life time. These properties can be expressed in terms of a quasiparticle self-energy $\Sigma = \Sigma' + i\Sigma''$, an analytical function the real and imaginary parts of which are related by the KK transform (the dispersion relation) \cite{Landau}. Neglecting the effects of the energy and momentum resolutions as well as the influence of matrix elements \cite{BorisenkoPRB2001}, one can take $I \propto A(\omega,\mathbf{k})$, where $\omega = E_k - h\nu + \phi$ is the energy of the electron in the crystal with respect to the Fermi level, $h\nu$ is the excitation energy and $\phi$ is the work function. In turn, the spectral function can be formulated in terms of the self-energy:
\begin{eqnarray}\label{E1}
A(\omega, \mathbf{k}) = -\frac{1}{\pi}\frac{\Sigma''(\omega)}{(\omega - \varepsilon(\mathbf{k}) - \Sigma'(\omega))^2 + \Sigma''(\omega)^2},
\end{eqnarray}
where $\varepsilon(\mathbf{k})$ is the bare band dispersion.

In case there is no interaction, i.e.~electronic excitations can live forever, the spectral function is a delta function with the pole $\omega - \varepsilon(\mathbf{k}_{\omega}) = 0$ and, e.g.~for the nodal direction, can be represented by the solid line in Fig.~\ref{Fig1}. When interactions are present, the self-energy leads to a shifting and broadening of the non-interacting spectral function. The resulting picture is essentially that which is measured in ARPES (the blurred region in Fig.~\ref{Fig1} illustrates this). If one neglects the momentum dependence of the self-energy, then, from Eq.~(\ref{E1}), the momentum distribution curves (MDC$(k) = A(k)_{\omega = const}$) have maxima at $k_m(\omega)$ determined by $\omega - \varepsilon(k_m) - \Sigma'(\omega) = 0$ for a given $\omega$. In other words, $\Sigma'(\omega) = \omega - \varepsilon(k_m)$, is that which is illustrated in Fig.~\ref{Fig1} by the double headed arrow. In the region where the bare dispersion can be considered as linear ($\varepsilon = v_F k$), assuming weak $\mathbf{k}$-dependence of $\Sigma''$ along a cut perpendicular to the Fermi surface \cite{BareApp}, the MDCs exhibit a Lorentzian lineshape \cite{KaminskiPRL01} with the half width at half maximum $W$ and $\Sigma''(\omega) = -v_F W(\omega)$. 

Thus, the determination of both the real and imaginary parts of the self-energy requires the knowledge of the bare dispersion (or, in the vicinity to $E_F$, an "energy scale", e.g., $v_F$ \cite{KordyukPRB2003}). The KK transformation gives an additional equation which relates these functions: 
\begin{eqnarray}\label{KK}
\Sigma'(\omega) = \frac{1}{\pi}\,\,PV\int_{-\infty}^\infty{\frac{\Sigma''(x)}{x - \omega}\,dx}, 
\end{eqnarray}
where $PV$ denotes the Cauchy principal value. This opens the way to extract all desired quantities from the experiment. 

In the vicinity of the Fermi level, one can determine the coupling strength as $\lambda = - (d\Sigma'(\omega)/d\omega)_{\omega=0} = v_F/v_R -1$, where $v_R = (dk_m/d\omega)_{\omega=0}^{-1}$ is the renormalized Fermi velocity. Differentiating the KK relation,
\begin{eqnarray}\label{E2}
\frac{d\Sigma'(\omega)}{d\omega} = \frac{1}{\pi}\,\,PV\int_{-\infty}^\infty{\frac{\Sigma''(x)-\Sigma''(\omega)}{(x - \omega)^2}\,dx},
\end{eqnarray}
for an even $\Sigma''(\omega)$,
\begin{eqnarray}\label{E3}
\lambda = \frac{-2}{\pi}\,\,PV\int_0^\infty{\frac{\Sigma''(\omega) - \Sigma''(0)}{\omega^2}\,d\omega} \equiv -\mathbf{D} \Sigma''.
\end{eqnarray}
Using the above definition of the $\mathbf{D}$ operator, $v_F^{-1} = v_R^{-1} - \mathbf{D} W$, or $1 + \lambda = 1/Z $, where
\begin{eqnarray}\label{E4}
Z = 1 - v_R \mathbf{D} W
\end{eqnarray}
is the coherence factor ($0 < Z < 1$).

In case $W(\omega)$ decays to zero or saturates on the scale covered by experiment, as it is expected for the scattering by phonons \cite{VallaPRL99}, the parameters $v_F$, $\lambda$ or $Z$ can be easily determined from the experimental values of $v_R$ and $\mathbf{D}W$. In cuprates, however, the MDC width $W$ along the nodal direction does not decrease or even saturate in the whole experimentally accessible energy region (up to $\omega_m =$ 0.5 eV) and one can make only a rough estimation expanding $\mathbf{D}W = \mathbf{D}_0^{\omega_m}W_{exp} + \mathbf{D}_{\omega_m}^\infty W_{mod}$, where $W_{exp}$ is the experimentally determined function of $\omega$, and $W_{mod}$ is a model function which depends on both the high energy cut-off, $|\omega_c| > |\omega_m|$, above which $\Sigma''(\omega)$ starts to decrease or saturate and a model for these high energy tails. For a simple estimation one can take $W_{exp} = \alpha \omega^2$ and $W_{mod} = \alpha \omega_m^2$ which gives $\mathbf{D}_0^{\omega_m}W_{exp} = \mathbf{D}_{\omega_m}^\infty W_{mod}$, demonstrating that the contribution of unknown high energy tails can be essential. 

Since without the knowledge of $W_{mod}(\omega)$ it is not possible to determine precisely the energy scale of the bare band or interactions from just the low binding energy part of ARPES spectra, in the following we examine a wider energy range of the ARPES data in order to solve the problem. In this energy range a deviation of the bare dispersion from a line should be taken into account. In Ref.~\onlinecite{KordyukPRB2003} we have shown that the shape of the bare band can be determined from the shape of the Fermi surface through the tight binding (TB) fitting procedure but the energy scale must be determined independently. 

Along the nodal direction the TB band in the occupied part can be well approximated by a simple parabola $\varepsilon(k) = \omega_0(1-k^2/k_F^2)$, where the energy scale parameter is the bottom of the bare band, $\omega_0$, or the bare Fermi velocity, $v_F = -2 \omega_0/k_F$. Then, using this dispersion in (\ref{E1}), one can express the complex self-energy function in terms of two experimental functions $k_m(\omega)$ and $W(\omega)$:
\begin{eqnarray}
\label{E5}\Sigma'(\omega) = \frac{v_F}{2 k_F} (k_m^2(\omega)-k_F^2) + \omega, \\
\label{E6}\Sigma''(\omega) = -\frac{v_F}{k_F} W(\omega) \sqrt{k_m^2(\omega)-W^2(\omega)}.
\end{eqnarray}
The KK transform $\Sigma'' \rightarrow \Sigma'$ completes the system but adds new parameters which describe the high energy tails of $\Sigma''(\omega)$.

\begin{figure}[t!]
\includegraphics[width=8.2cm]{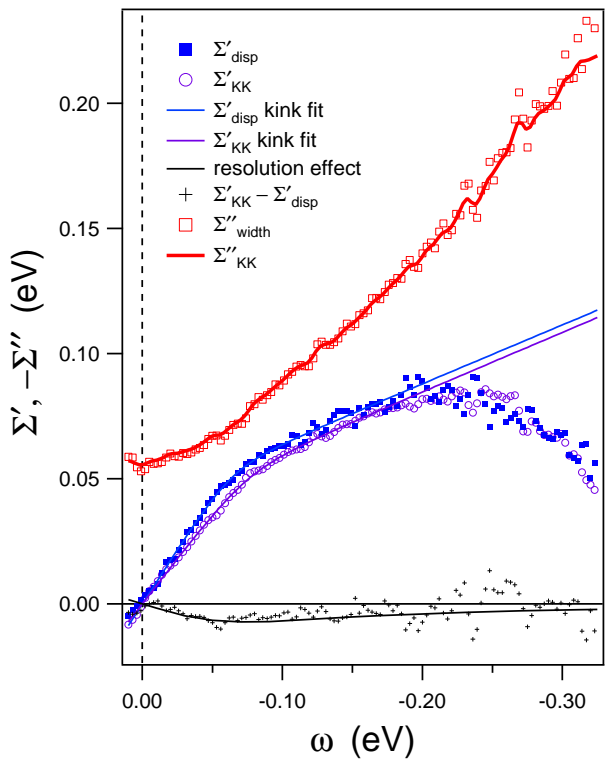}%
\caption{\label{Fig2} Real and imaginary parts of the self-energy extracted from the experiment with the described procedure. A complete coincidence between the corresponding parts of the self-energy calculated from the two different experimental functions, the MDC dispersion and MDC width, demonstrates the full self-consistency of the ARPES data treated within the self-energy approach.}
\end{figure}

As an example, we use the experimental data from an underdoped Bi(Pb)-2212 ($T_c =$ 77 K) measured along the nodal direction at 130 K with 27 eV excitation energy, at which only the bonding band is visible \cite{NBS}. The experimental details can be found elsewhere \cite{KimPRL2003, Drop}. 

Fig.~\ref{Fig2} shows the results of the procedure we describe. The real part of the self-energy is determined in two ways: (i) $\Sigma'_\textrm{disp}$, by Eq.~(\ref{E5}) with $v_F$ as a parameter; (ii) by Eq.~(\ref{E6}) and subsequent KK transform of symmetrized $\Sigma''_\textrm{width} \rightarrow \Sigma'_\textrm{KK}$ with additional parameters which describe the $\Sigma''(\omega)$ tails, viz.~a cut-off energy, $\omega_c$, at which the tail is attached and a model of its decay (see \cite{BareApp} for details). If the self-energy we are trying to extract satisfies the KK relation, $\Sigma'_\textrm{disp}$ and $\Sigma'_\textrm{KK}$ should be identical within experimental uncertainty which sets the space of the parameters in question.

In the experimental energy range of 350 meV we have found that $v_F$ is restricted by such a procedure in contrast to $\omega_c$ and to the model for the high energy tails, for which only a relation can be obtained. In practice, we checked several models where $\Sigma''_\textrm{mod}(\omega)$ saturates to some maximum value, slowly decays or drops to zero above $\omega_c$, and for each model the fitting of $\Sigma'_\textrm{KK}$ to $\Sigma'_\textrm{disp}$ results in the same $v_F$ (or $\omega_0$) but different $\omega_c$ (for details see Fig.~2 in \cite{BareApp} and related discussion). 

At this stage we have found that although the fitting procedure is working well, i.e.~the standard deviation has a minimum as a function of ($\omega_0$, $\omega_c$), and the resulting curves $\Sigma'_\textrm{disp}$ and $\Sigma'_\textrm{KK}$ almost coincide (not shown), there are still deviations between them which exceed the experimental error. These deviations can be expressed for these two curves as a nondecreasing difference in the slope ratios below and above the 70 meV kink. To visualize this we fit the $\Sigma'(\omega)$ functions in the range of $|\omega| <$ 170 meV to $\Sigma'_\textrm{low}(\omega) = -\lambda \omega -  \Delta\lambda (\omega - \omega_k) \{\arctan[\omega_k / \delta] - \arctan[(\omega - \omega_k)/\delta]\} /\pi$ (see Fig.~\ref{Fig2}), where $\omega_k \approx -70$ meV is a kink energy and $\delta \approx$ 30 meV is a kink width \cite{kwidth}. For $\delta \ll |\omega_k|$, $\lambda$ and $\lambda_h = \lambda - \Delta\lambda$ reflect the slopes below ($|\omega| < |\omega_k|$) and above the kink energy respectively, and it is the difference in $\lambda/\lambda_h$ between $\Sigma'_\textrm{disp}$ and $\Sigma'_\textrm{KK}$ which is irreducible by varying the parameters $\omega_0$, $\omega_c$ and the high energy tails. 

The origin of this difference is related to resolution which mainly effects the low energy part of $\Sigma'_\textrm{KK}(\omega)$ through $W(\omega)$. Therefore, high energy slopes $\lambda_h$ and high energy tails which are resolution independent should coincide for $\Sigma'_\textrm{disp}$ and $\Sigma'_\textrm{KK}$ after the fitting procedure is performed. This appeared to be true and the final result is shown in Fig.~\ref{Fig2}. It is remarkable that the difference $\Sigma'_\textrm{KK} - \Sigma'_\textrm{disp}$ completely coincides with $R'(\omega)$, the contribution from the overall resolution $R =$ 15 meV, calculated as $R' = $ KK~$R''$, where $R''(\omega) = \sqrt{R^2 + [\Sigma''(\omega)-\Sigma''(0)]^2} - [\Sigma''(\omega)-\Sigma''(0)]$ is the contribution of the overall resolution to the scattering rate. Such a complete coincidence substantiates that the self-energy constructed using Eqs.~(\ref{E5}) and (\ref{E6}) is self-consistent within the experimental accuracy currently available with ARPES. In order to check the correctness of the KK numerics, we also plot $\Sigma''_{KK}(\omega)$ function which is obtained by back KK-transform of $\Sigma'_{KK}(\omega)$.

\begin{figure}[b!]
\includegraphics[width=8.2cm]{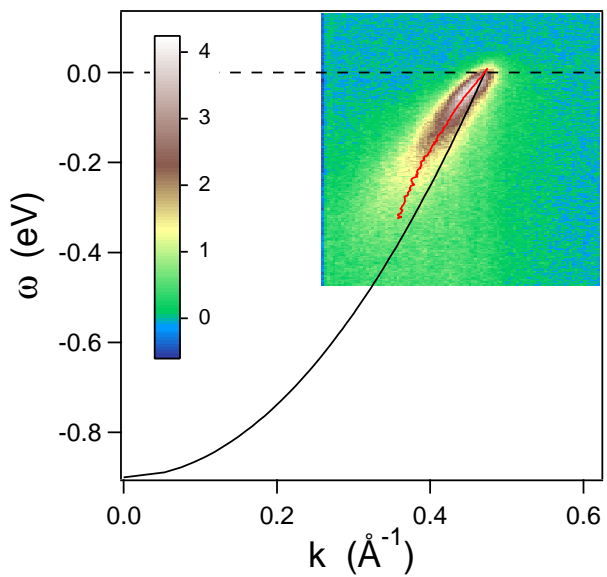}%
\caption{\label{Fig3} The bare band dispersion along the nodal direction of an underdoped Bi(Pb)-2212 (black solid line) on top of its spectral weight at 130 K measured by ARPES. MDC (or renormalized) dispersion shown by solid red line.}
\end{figure}

For underdoped Bi(Pb)-2212, for which $k_F =$ 0.471 \AA$^{-1}$ and $v_R = 2.04 \pm 0.05$ eV\AA, we have obtained for the bare band dispersion along the nodal direction $v_F = 3.82 \pm 0.17$ eV\AA~($\omega_0 = - 0.90 \pm 0.04$ eV). This bare band is shown in Fig.~\ref{Fig3} on top of the electron spectral weight as measured by ARPES. The interaction parameters thus are the following: $\lambda = 0.87 \pm 0.12$ and $\lambda_h = 0.24 \pm 0.03$. As we discussed above, the energy cut-off, $\omega_c$, depends on the model used for the high energy tails of $\Sigma''(\omega)$ but we can state that $|\omega_c| \approx |\omega_0|/2$.

The linear behavior of $\Sigma'(\omega)$ over a wide energy range $|\omega| < |\omega_k|$ indicates, using the criterion $\lim_{\omega \to 0}\Sigma''(\omega)/\omega = 0$, the existence of well defined quasiparticles in the pseudogap state: for the underdoped Bi(Pb)-2212 at 130 K the coherence factor $Z = 0.54 \pm 0.03$. The offset of $\Sigma''(\omega)$ not only comes from finite resolution but also finite temperature and scattering on impurities, which are mostly energy independent \cite{Drop} and do not contribute to the slope of $\Sigma'(\omega)$ and, therefore, to the coherence factor.

In \cite{KoitzschPRB2004} we have noticed that the scattering rate at room temperature looks more linear for underdoped samples than for overdoped ones that is in favour of the marginal Fermi liquid model (MFL) \cite{VarmaPRL89}. It is important to stress that $\Sigma'(\omega)$ determined with better accuracy exhibits a linear behavior below and above the kink energy $\omega_k$ (see Fig.~\ref{Fig2}) which is now difficult to reconcile with the MFL model: as far as a slope in $\Sigma'(\omega)$, according to Eq.~(\ref{E2}), is mainly determined by the coefficient at $(\omega - \omega_x)^2$ term in the expansion of $\Sigma''(\omega)$ around $\omega_x$, the straight sections on $\Sigma'(\omega)$ imply the regions where $\Sigma''(\omega)$ is precisely parabolic (exhibits constant curvature over some finite energy regions). 

Another point arises as a consequence of the tight correlation between $\Sigma'$ and $\Sigma''$. Recently we have shown \cite{Drop} that two different channels can be distinguished in the scattering rate: the doping independent Auger like decay which originates from the electron-electron Coulomb interaction and the doping dependent one which can be naturally assosiated with spin excitations. While such a decomposition of the scattering rate into two channels seems to be becoming commonly accepted \cite{ZhouXXX04}, there is still a controversy about the origin of the doping dependent one. The present analysis shows that regardless of the nature of this channel, its doping and temperature dependence should appear in the doping and temperature dependence of $\Sigma'$ and, consequently, of the renormalized dispersion, although it is clear that the variations in the latter should be marginal. This casts doubt on the apparent universality of the nodal Fermi velocity \cite{ZhouNature03}, viz. that under close scrutiny it will not hold.

In conclusion, we demonstrate for the first time the full self-consistency of the data obtained using angle resolved photoemission and treated within the self-energy approach. This self-consistency is demonstrated on an underdoped Bi-2212 sample ($T_c =$ 77 K) in the pseudogap state. The extracted bare band dispersion is in good agreement with the band structure calculations and allows one to quantify the self-energy of the electronic excitations in the real energy scale. The accurately determined real and imaginary parts of the self-energy prove the existence of well defined quasiparticles along the nodal direction even in the pseudogap state of Bi-2212.

The project is part of the Forschergruppe FOR538 and is supported by the DFG under grants number KN393/4 and 436UKR17/10/04 and by the Swiss National Science Foundation and its NCCR Network "Materials with Novel Electronic Properties".


\begin{thebibliography}{}
\bibitem{kwidth} Half width at quarter maximum of $-d^2\Sigma'_\textrm{low}(\omega)/d\omega^2$. Here we introduce an analytical function in order to quantify a peculiarity in the experimental dispersion known as a "kink".
\end{thebibliography}
\end{document}